\documentstyle{article}

\begin{document}

\begin{center}
{\Large\bf 2D Lattice of coupled Sinai billiards:\\
metal or insulator at $g\ll 1$?}

\medskip
{\large M.V.Budantsev$^{1,2}$, Z.D.Kvon$^1$, A.G.Pogosov$^1$,
G.M.Gusev$^3$,\\
J.C.Portal$^2$, D.K.Maude$^2$, N.T.Moshegov$^1$, A.I.Toropov$^1$}

\medskip\em
$^1$Institute of Semiconductor Physics, pr. Lavrentyeva,
13, Novosibirsk 630090, Russia

\medskip
$^2$Grenoble High Magnetic Field Laboratory, MPI-FKF and CNRS,
F-38042 Grenoble, France

\medskip
$^3$Instituto de Fsica da Universidade de So Paulo, CP 66318, CEP
05315-970, Brasil
\end{center}

\begin{abstract}
We investigate the transport in a two-dimensional (2D) lattice of coupled
Sinai billiards fabricated on the basis of a high-mobility 2D electron gas
in GaAs/AlGaAs heterojunction. For the states with low reduced
conductivity $g\ll 1$ an anomalously weak temperature dependence of $g$
was found. The large negative magnetoresistance described by the Lorentz
line-shape of the width corresponding to the half magnetic flux quantum
through the area of the billiard is observed. In going from $g>1$ to
$g\ll 1$ it strongly increases. The Shubnikov-de Haas oscillations and
commensurability magnetoresistance peak are preserved at $g\ll 1$. The
data suggest that the system studied behaves more like a metal than an
insulator at $g\ll 1$ and is not described by the generally accepted
picture of Anderson localization.
\end{abstract}

Quantum and classical transport in systems with dynamic chaos has
been intensively studied during last few years, since the successes of
modern semiconductor technology has made it possible to obtain various
experimental realizations of such systems with electron billiards as an
example. At the present time two varieties of these systems are studied.
The first one is unit billiards (regular or chaotic Bunimovich or Sinai
billiards) [1--5], while the second one is macroscopic two-dimensional
(antidot lattice) [6--10] or one-dimensional [11] Sinai billiards. A
number of interesting phenomena resulting from the classical and quantum
chaotic electron dynamics was found in these billiards (mesoscopic
conductance fluctuations, commensurability magnetoresistance oscillations,
weak localization effects, statistical and parametrical correlations of
Coulomb blockade peaks).

However up to now, there have been practically no investigations of
systems, in which separate electron billiards with reduced
conductivity $g=\sigma_{xx}/(e^2/h)\gg 1$ and the size $L<l$ ($l$ is the
mean free path) compose a lattice with the coupling between them being
weak in the sense that the conductivity of the lattice itself is low $g\ll
1$. The study of such lattice is of interest mainly due to the fact that,
on the one hand, this system should have some properties of a unit
billiard (because the coupling is weak), and on the other hand, since
the billiards form a regular structure, the systems may exhibit properties
caused by that structure.

We report for the first time the results of experimental investigation of
this new system with dynamical chaos. The system was built on
the basis of a two-dimensional (2D) lattice of closely situated antidots
fabricated from a high mobility 2D electron gas in GaAs/AlGaAs
heterojunction with a metallic gate evaporated on the top, that permitted us
to control the conductivity of the structure in a sufficiently wide range
from $g=0.01$ to $g=2$.

The square lattice of antidots was fabricated on the basis of a
two-dimensional electron gas with electron density $N_S=(2-3)\times
10^{11}$~cm$^{-2}$ and mobility $\mu=(3-8)\times 10^5$~cm$^2$/V~s
corresponding to the mean free path $l=(3-6)~\mu$m by means of electron
lithography and subsequent plasma etching. Then the NiAu or TiAu gate was
evaporated on the top of the device. We investigated three samples with
the lattice period $d=0.6~\mu$m and three ones with $d=0.7~\mu$m. One of
the samples had no metallic gate, and its conductivity was controlled using
illumination by LED. The lithographic size of antidots $a=0.2~\mu$m was
the same for all samples. However, due to the depletion layers the actual
size was larger, being approximately equal to the lattice period even
before the gate evaporation. The experimental sample was made up of two
Hall bars of the length of 100~$\mu$m and the width of 50~$\mu$m. The
lattice of antidots was introduced into one of the bars.  Measurement were
carried out in the temperature range 50~mK--1~K in the magnetic field up
to 2~T using ordinary four-terminal scheme at the frequency 6~Hz and with
the current 0.01--0.1~nA in order to exclude heating effects.

Fig.~1(a) shows the set of temperature dependences of the conductivity
$g(T)$ for the sample AG219 with the period $d=0.6~\mu$m ($\mu=7\times
10^5$~cm$^2$/V~s at $N_S=2\times 10^{11}$~cm$^{-2}$ in the unpatterned part
of the sample) for different values of the gate voltage. It is seen that
for $g>1$ conductivity is practically temperature independent. More
exactly, weak logarithmic decrease of $g$ is observed typical for the weak
localization effects. It becomes more noticeable for lower values of
conductivity, but it still remains weak even for $g\ll 1$, and is well
described by the power law dependence $g(T)\propto T^a$ with $a<1$ for all
of the tested samples. Specifically, for the dependences of Fig.~1(a)
$a=0.1-0.27$ at $g\ll 1$. It should be mentioned that the value of $a$
is larger for the samples with lower mobility of the initial
2D electron gas. As an example, the same set of
dependences $g(T)$ for sample AG35 with the period $d=0.7~\mu$m
($\mu=3\times 10^5$~cm$^2$/V~s at $N_S=2\times 10^{11}$~cm${-2}$) is
presented in the Fig.~1(b). It is seen that they are characterized by
two times higher magnitude of $a$ for the same values of $g$. Hence it
may be inferred that the enhancement of fluctuation potential leads to
stronger temperature dependence. The behavior of $g(T)$ described above
is significantly different from that for the unpatterned 2D electron gas
both in silicon MOS-transistors [12] and in AlGaAs/GaAs heterojunction
[13], as well as for antidot lattices with short period [14,15]. In all
of these cases at $g\sim 1$ the transition from the weak logarithmic
dependence (weak localization regime) to the strong exponential one
(strong localization regime) is observed. In our case the weak logarithmic
decrease of $g$ (for $g>1$) is followed by a weak power law, that has not
been observed in other 2D systems before.

We now turn to description of the influence of magnetic field. It is
well known that in AlGaAs/GaAs heterojunctions at $g<1$ under the
influence of magnetic field a transition from an insulator to the
quantum-Hall-liquid is observed. This transition is characterized by the
critical point $B_c$ and $g_c\approx 0.5-1$ [13]. A similar transition was
recently observed in the triangular antidot lattice with a small period
$d=0.2~\mu$m [15]. Our samples exhibit a radically different picture. It
is seen from Fig.~2 that for all values of $g(B=0)$ in the magnetic fields
about $B\approx 1$~T the transition takes place from  weak power-law
dependence $g(T)$ to no temperature dependence at all. Moreover, this
transition is of different kind, for there is no critical point, and the
metallic behavior extends for $g\ll 1$. Fig.~2(a) also shows an
interesting picture in weak magnetic field.
For all values of $g$, the negative
magnetoresistance (NMR) is observed for $B<0.05$~T, followed by a peak
at $B\approx 0.2$~T corresponding to the condition $2R_c=d$. This peak is
well-known for the antidot lattices at $g>1$ and originates from the
so-called pin-ball trajectory that surrounds an antidot not colliding with
it. It is seen from Fig.~2(b) that the second commensurate peak is
observed at $g\approx 1$. The second peak corresponds to the condition
$2R_c=(\sqrt{2}-1)d$. We have recently established that this peak is
associated with the non-colliding trajectory inside the billiard [16].
The positions of the commensurability peaks shows that we really deal with
the lattice of closely situated antidots with $d\approx a$ and $d\gg d-a$.
The Sinai billiards between the antidots have the area $S=d^2(1-\pi/4)$
and contain a large number of electrons $N\gg 1$.  In our case we have
correspondingly $S=0.5~\mu$m$^2$, $N\approx 70$ for $d=0.7~\mu$m, and
$S=0.36~\mu$m$^2$, $N\approx 50$ for $d=0.6~\mu$m. It is also important
that both the main commensurability peak and NMR are conserved at the
transition from $g>1$ to $g\ll 1$. The peak position is slightly shifted
towards lower magnetic fields with decreasing $g$. This indicates
a small change of electron density (from $1.4\times 10^{11}$ to
$0.9\times 10^{11}$~cm$^{-2}$) in the billiard within the whole range of
$g$.

The behavior of NMR is shown in Fig.~3 in more detail. It is
characterized by two distinguishing features: (i) for all states with
$0.05<g<2$ NMR is cut off at the same magnitude of magnetic
field $B\approx 0.05$~T; (ii) NMR noticeably increases with decreasing $g$
(Fig.~3(a) shows that at $g=0.05$ it reaches a considerable value about
40\%), and its temperature dependence becomes stronger. For the states
with the highest resistivity it was more stronger than
$g(T)$).  For $g>1$ NMR can be attributed to the effects of weak
localization in open chaotic billiards [2], because it has relatively
small amplitude and Lorentzian line-shape. The behavior of NMR for $g\ll 1$
is surprising. It increases by an order of magnitude reaching a
considerable value comparable to the total resistance of the sample, while
the line-shape of NMR is described by a Lorentz curve of the same width.
The width is equal to $\Delta B_1=27\pm 2$~mT for the sample AG219.
It corresponds to a half magnetic flux quantum through the area of the
billiard that is equal to $d^2(1-\pi/4)$.  The fact that the width is
determined by the magnetic flux quantum is well seen from the comparison
of NMR for the samples with two different periods 0.6~$\mu$m
and 0.7~$\mu$m.  As it is clearly seen from the
Fig.~3(c) the width of NMR curve for the period 0.7~$\mu$m equals $\Delta
B_2=20\pm 2$~mT, that is $\Delta B_1/\Delta B_2=(0.7)^2/(0.6)^2$.
Thus at $g\ll 1$ we observe NMR wich is very similar to weak localization
NMR in chaotic open billiards [2]. But the value of NMR is much larger. It
is necessary to add that the following two conditions should be satisfied
in order to observe the effects described above: (i) the influence of the
fluctuation potential should be as low as possible, and (ii) the area of
billiard should be relatively large.  That is because one or both these
conditions were not met that these effects were not observed in [13,14,17]
(except for the commensurability peak observed in [14] at $g\ll 1$, that
might still have been of different nature, for the temperature dependence
was exponentially strong).

Let us discuss the results obtained. First turn to the temperature
dependence at the transition from $g>1$ to $g\ll 1$. It differs from
the accepted picture of the metal-insulator transition in 2D and 3D
electron systems. This picture is based on the concept of Anderson
localization of electrons, be it the model of minimal metallic
conductivity (MMC) or the scaling theory (ST) [18]. According to it at
$k_Fl\sim 1$ or $g\sim 1$ in a macroscopic 2D or 3D system the transition
should occur from the metallic behavior (as in MMC model) or from the weak
localization (as in ST) to the strong localization characterized by the
exponential temperature dependence of the activation type
(hopping conductivity) or of the Mott type (variable range hopping
conductivity).  In real systems the transition can be quite complicated,
but for $g\ll 1$ the state with hopping conductivity is always realized
[12,13]. We know only one work [19] in which the linear temperature
dependence for $g\ll 1$ was observed in thin In$_2$O$_3$ films. Recently
a similar system was considered theoretically in [20] within
the model of metal grains with $g\gg 1$ coupled by tunneling in a way that
the conductivity of the macroscopic sample was low $g\ll 1$.
At $g\ll 1$, due to inelastic electron-electron scattering the linear
dependence $g(T)$ was obtained in [20] in a wide temperature range. The
dependence changes to the exponential one at $T\ll e^2/C$, where $C$
is the capacitance of a metal grain. Our results are not in
agreement with the predictions of this theory. Firstly, in our case
$g(T)$ is weaker than linear. Secondly, it is observed at $T\ll
e^2/C$, because the Coulomb energy for our samples
$e^2/2C\approx$15--25~K.  This means that the low conductivity of the
lattice of the Sinai billiards can not be explained by the model of metal
lakes coupled by weak tunnelling junctions. The effects of Coulomb
blockade are not manifested in the experiment, because one observe no
features in $g(V_g)$ dependence. So we have to assume that at $g\ll 1$ the
coupling between the billiards is stronger than that provided by
tunneling. The behavior of the lattice in the magnetic field supports this
assumption. In Fig.~2(a) one can see the Shubnikov--de Haas oscillations,
which give the electron density in the lattice saddle points connecting
the billiards. It coincides with the density determined from the Hall
effect that should give the concentration in these points [21]. The
electron density in these points weakly changes with the strong change of
$g$.  It decreases only by about 30\% while $g$ drops by a factor of 30,
and its magnitude, equaled to $4.1\times 10^{10}$~cm$^{-2}$, is only three
times less than $N_S$ inside the billiard even for the state with the
lowest value of $g$. This means that the Fermi level in the saddle point
lies several meV above the barrier at $g\ll 1$, and an electron should
ballistically move through the "bottle neck" to go from one billiard to
another. The behavior of commensurability peak and of NMR support this
picture. Hence, we should come to a paradoxical conclusion that 2D lattice
of Sinai billiards coupled via conducting "bottle-neck" can have very low
conductivity $g\ll 1$ but simultaneously exhibits the properties typical
for metallic ballistic systems rather then for the insulators. This
conclusion is in a drastic contradiction with the standard picture of
Anderson localization.  Let us discuss possible reasons for such a
situation and consider it first from the weak localization side. The
transition from the weak to strong localization, caused, for example, by
the decrease of temperature, should be accompanied by the increase of the
phase coherence length. Due to this increase, the number of localized
trajectories should increase until at $T=0$ all the trajectories become
localized.  In the system investigated this increase can be at least
hampered, because an electron can loose the phase coherence inside a
billiard before it leaves for the other billiard through the
"bottle-neck". A simple estimation gives for the dwell-time of the
electron inside a billiard $\tau=10^{-8}$~s.  The collimation effects can
only increase this time. The estimation shows that $\tau$ can be larger
than the time of phase coherence, which is of the order of
$\tau_\varphi=10^{-9}$~s at 40~mK. This means that even at the lowest
available temperature an electron can loose the phase coherence on the
length scale $L=d$. Then the resistance is the classical sum of the
resistances of individual billiards, which can yield any value of $g$. The
situation discussed is to some extent similar to that considered in [20].
In contrast to [20] in our system the quantum dots are separated by the
conducting "bottle-neck" instead of the tunnel barrier as in [20].
Obviously, this should lead to the weaker temperature dependence of the
conductivity, that is just observed in our experiments.  Nevertheless, it
is not clear what happens at $T\rightarrow 0$, because the Coulomb
blockade effects are not observed in our case. The description of the
system from the strong localization side is complicated, because one can
not start from the ground state of electron in the well, for it represents
a multilevel system with the large ($\sim 100$) number of electrons. In
any case the description of 2D lattice of coupled Sinai billiards and the
phenomena described in the present work presents a challenge to the modern
theory of quantum transport in the condensed matter.

     In conclusion, we investigate a new type of dynamic chaos
system --- 2D lattice of coupled Sinai billiards. We find that
its transport properties are  very unusual. They characterize the system
as a metal rather than an insulator at $g\ll 1$. The whole set of
experimental data (temperature dependence of conductivity, NMR, and
commensurability peak of magnetoresistance) is in contradiction with the
standard picture of Anderson localization, and testifies that the system
possesses quite unusual transport properties the description of which
requires new theoretical approaches.

	Note added. Since the completion of this paper, D.P.Druist et al [22]
have reported the observation of metallic behavior at $g\ll 1$ in a
completely different system --- a layered three-dimensional semiconductor
structure --- under the conditions of IQHE.

     We would like to thank M.Entin, V.Falko, A.Charplik, and E.Baskin for
useful discussion and V.Alperovich for reading the manuscript.
M.V.Budantsev acknowledges "Mission scientifique" from French Embassy in
Moscow for the support. This work was supported by RFFI through Grant No
96-02-287 and by NATO Linkage through grant No HTECH.LG.971304.

\newpage
\noindent
{\Large\bf Figure captions}

\bigskip
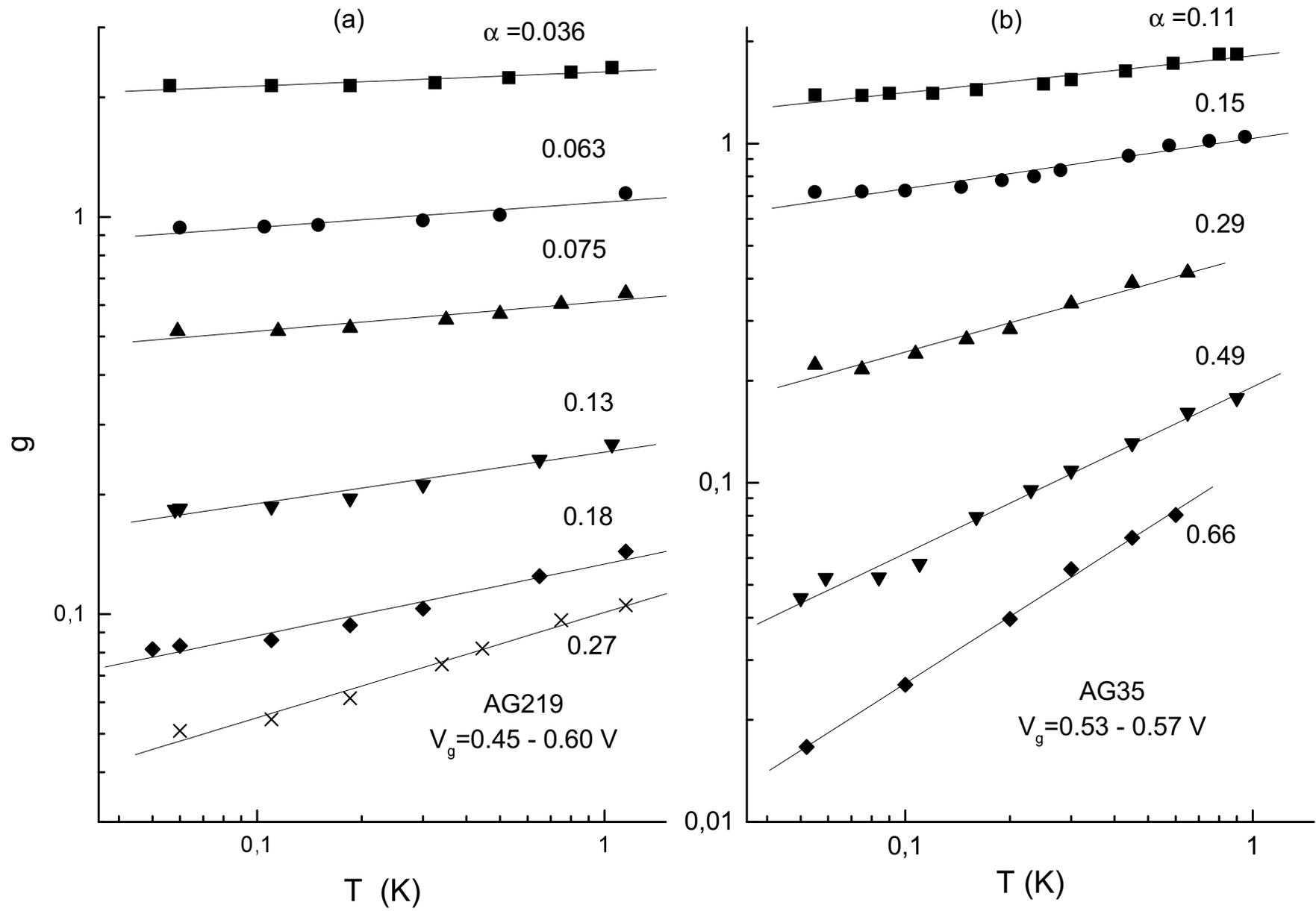
{\bf Fig. 1}\\
Temperature dependences of the conductivity at the transition
from $g>1$ to $g\ll 1$: (a) sample AG219, (b) sample AG35. Solid lines are
$g \sim T^a$.

\bigskip
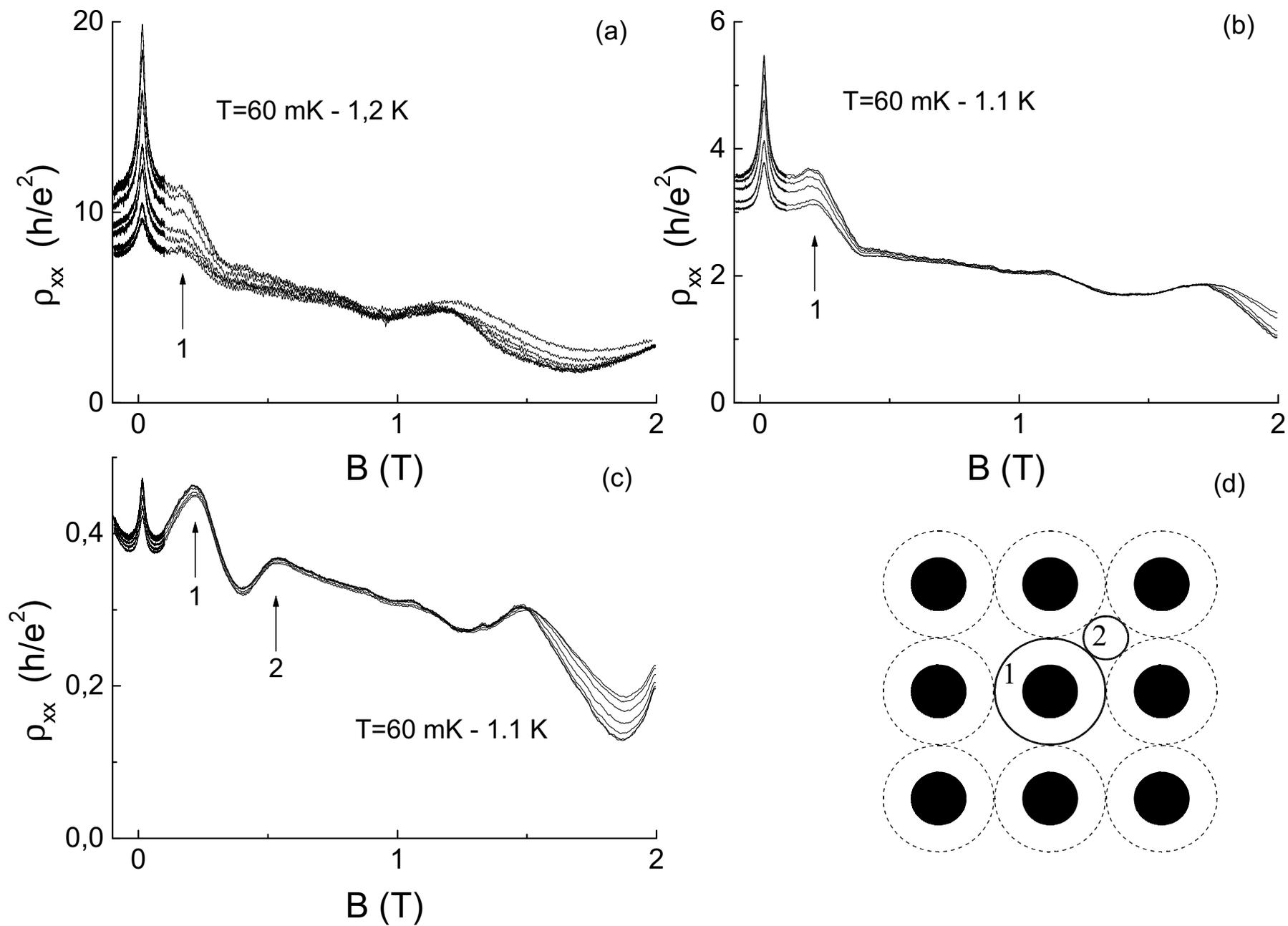
{\bf Fig. 2}\\
Magnetoresisance (MR) traces for different values of the
conductivity and at different temperatures (sample AG219): (a) $g=0.05$,
(b) $g=0.18$, (c) $g=2.2$. (d) Schematic view of antidot lattice (black
points show the etched regions, broken lines show the boundary of depletion
region, 1 - an electron trajectory around antidot, 2 - an electron
trajectory between antidots).

%\begin{figure}
%\caption{(a,b,c) A more detailed MR traces for weak magnetic field range
%at the same values of $g$ as in Fig.~2 at $T=60$~mK (sample AG219).  }
%\end{figure}

\bigskip
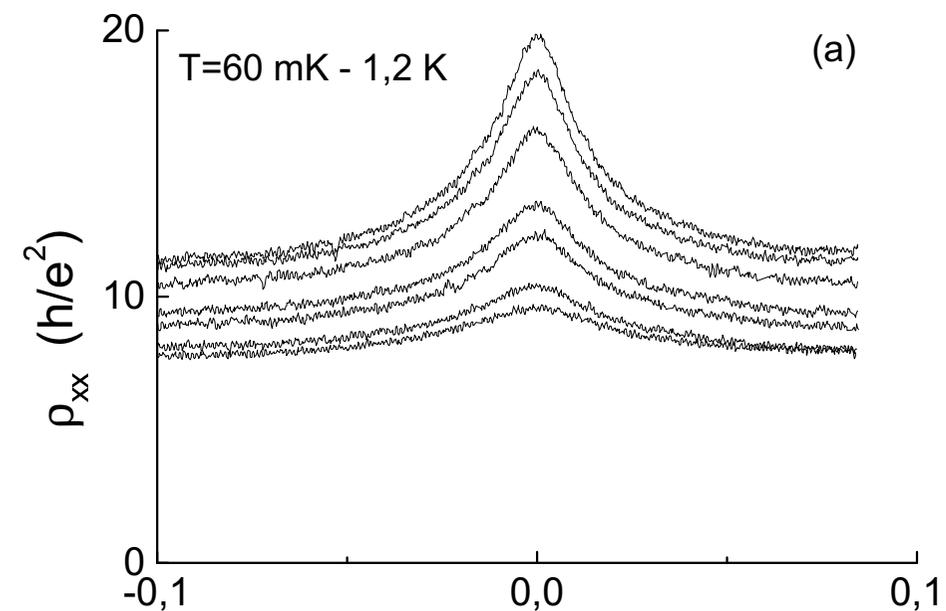
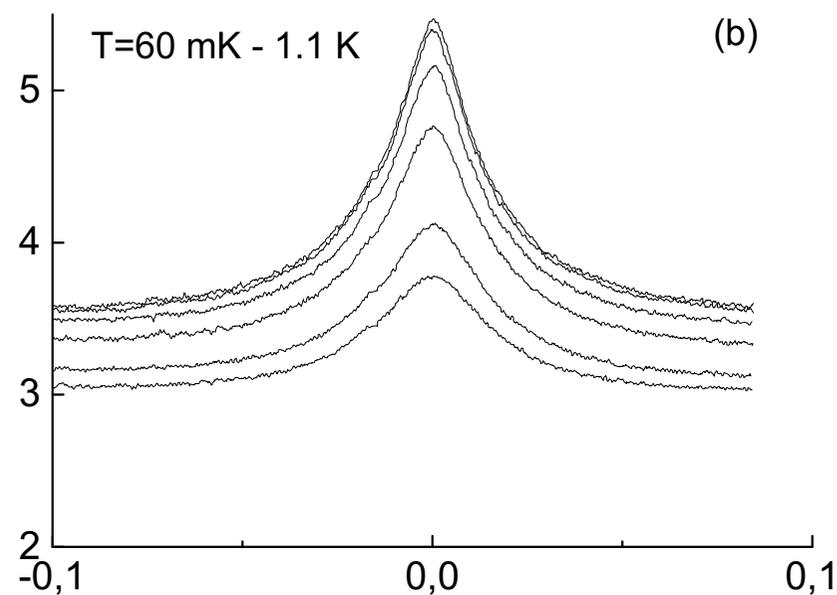
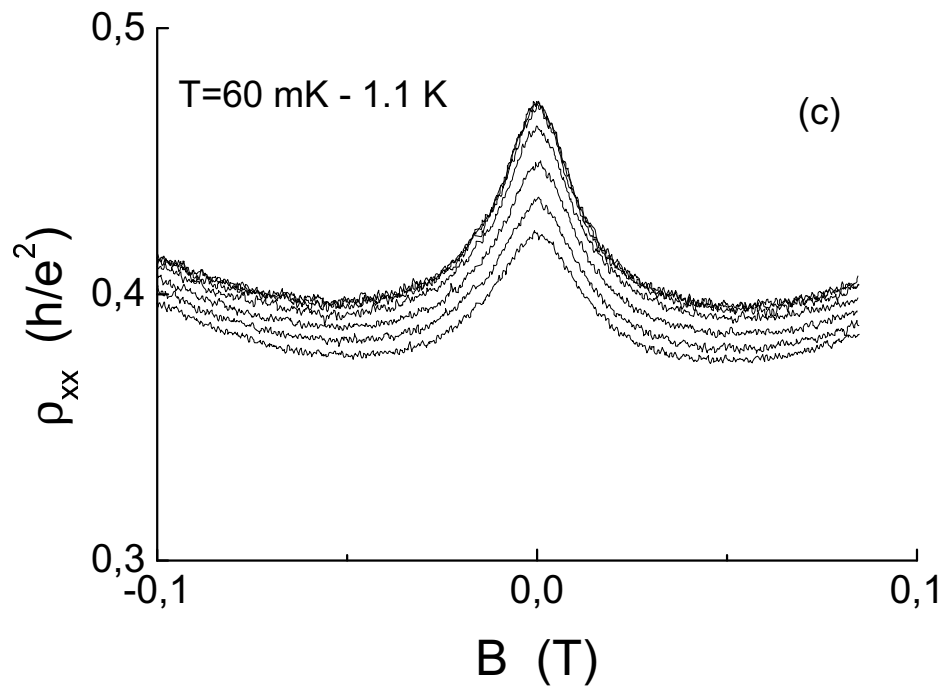
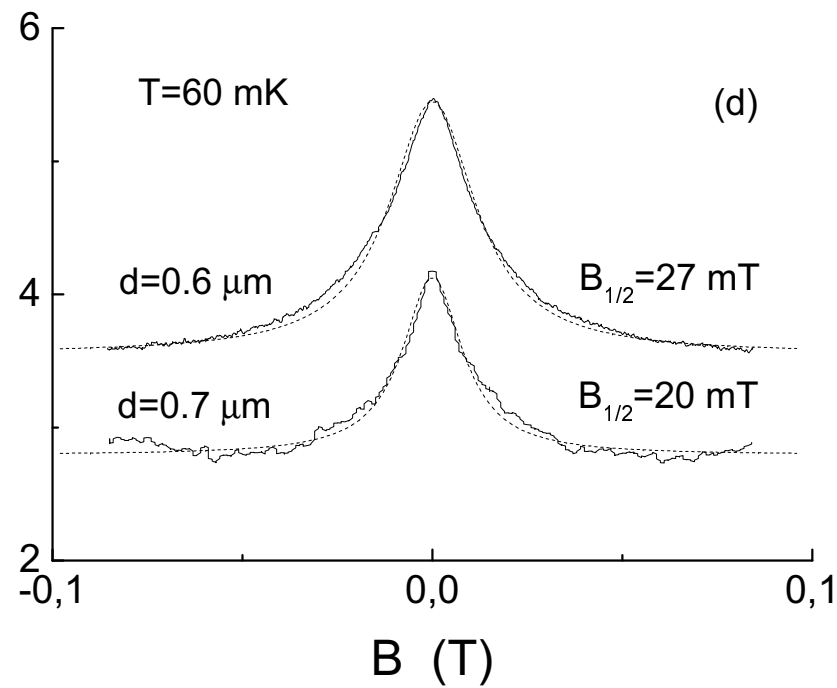
{\bf Fig. 3}\\
(a,b,c) NMR curves for the sample AG219. (d) Experimental and
calculated NMR curves for the antidot lattices with two different periods
$d=0.6~\mu$m (sample AG219) and $d=0.7~\mu$m (sample AG35). Solid lines are
the experimental curves, broken lines are the calculated Lorenz ones,
$B_{1/2}$ is the width of the Lorenz curves.

\end{document}